\documentclass[a4paper,12pt]{article}
\usepackage[ansinew]{inputenc}
\usepackage[T1]{fontenc}
\usepackage[english]{babel}
\usepackage{latexsym}
\usepackage{amsmath}
\usepackage{amssymb}
\usepackage[dvips]{graphicx}
\usepackage{epsfig}
\usepackage{subfigure}
\newcommand{\mysection}[1]    % Re-define the chaptering command to use
	{                   % THESE headers.
	\section{#1}
	 
	}

\newcommand{\mc}{\mathcal}

\newcommand{\eps}{\varepsilon}
\newcommand{\ol}{\overline}
\newcommand{\nl}{\newline}

\newcommand{\beq}{\begin{equation}}
\newcommand{\eeq}{\end{equation}}
\newcommand{\beqa}{\begin{eqnarray}}
\newcommand{\eeqa}{\end{eqnarray}}
\newcommand{\beqan}{\begin{eqnarray*}}
\newcommand{\eeqan}{\end{eqnarray*}}
\newcommand{\nn}{\nonumber}

\newcommand{\bc}{\begin{center}}
\newcommand{\ec}{\end{center}}
\newcommand{\Erf}{\textrm{Erf}}
\newcommand{\Rea}{\textrm{Re}}
\newcommand{\Ima}{\textrm{Im}}

\title{Numerical study of leptogenesis in a 5D split fermion model with bulk neutrinos}
\author{Heidi Kuismanen$^a$, Jukka Maalampi$^{b,c}$ {\small and} Iiro Vilja$^a$ \\ 
{\small $^a$ \textit{Department of Physics and Astronomy, University of Turku, 20014 Turku, Finland}}\\
{\small $^b$ \textit{Department of Physics, University of Jyväskylä, 40014 Jyväskylä, Finland}} \\ 
{\small $^c$ \textit{Helsinki Institute of Physics, FIN-00014 University of Helsinki, Finland}}} 
 
\date{\today}

\begin{document}

\maketitle

We study numerically a 5D hybrid model which incorporates a split fermion scenario and bulk neutrinos. We perform a Monte Carlo analysis of the model in order to find the regions in the parameter space allowing for realization of the leptogenesis.  We find that higher order Yukawa terms must be included in order the model to produce a CP violation and net baryon number sufficient for the creation of the observed baryon asymmetry of the Universe. 

\section{Introduction}
\setcounter{equation}{0}
\setcounter{footnote}{0}

The observed matter-antimatter asymmetry of the Universe suggests that the Standard Model (SM) of particle physics is not the final theory. From the premise that the Universe was matter-antimatter symmetric above the electroweak phase transition temperature $\mc O(100\ {\text{GeV}})$ SM predicts the ratio of present baryon number density over photon number density to be $n_B/n_\gamma \sim 10^{-18}$ \cite{smasymmetry}. One the other hand, the Wilkinson Microwave Anisotropy Probe (WMAP) \cite{wmap} measured the result 
\beqa
\frac{n_B}{n_\gamma}=(6.1\pm 0.3)\times 10^{-10},
\eeqa
which means there is a large discrepancy between observations and SM.

Baryogenesis via leptogenesis is considered to be one of the most appealing scenarios of producing matter-antimatter asymmetry in the Universe, \cite{Fukugita:1986hr, luty}. Contemporary models extend SM with heavy singlet neutrinos that undergo decay to leptons and antileptons, thus yielding net lepton number that is converted to net baryon number via sphaleron transitions during electroweak phase transition with the temperature $\sim \mc(100\ {\text {GeV}})$ \cite{krs}. These heavy neutrinos also account for the observed nonzero masses of SM neutrinos in the so called seesaw mechanism \cite{mohapatra}. 

There are various ways to extend SM to include heavy neutrinos. One approach is to consider $SO(10)$ models where the heavy neutrino mass is close to the GUT scale, while the other relies on superstring theories and introduces extra dimensions. In these extra dimensional models or brane models, our 4-dimensional Universe, the brane, is immersed into larger dimensional space-time, the bulk. The heavy neutrino can propagate in the bulk while SM particles cannot, \cite{arkanihamedrussell, dienesdudasgherghetta}. The advantage brought by brane models is the possibility to lower the fundamental scale of gravity many orders of magnitude below the effective gravity scale, $M_{Pl}$ and also account for the hierarchies within the SM fermion families \cite{arkanihamed}.      

In a hybrid model introduced in \cite{dienes} the bulk neutrino model is incorporated with the split fermion scenario \cite{thickbrane}. For every brane neutrino one can assign a bulk neutrino, which effectively brings flavor dependence to the model. This also results in a great number of undetermined parameters, which detracts from the predictive power of the model. Making the brane-bulk couplings flavor-neutral will not help either because then there is little leeway for the mass matrix to explain observed neutrino oscillations. In split fermion scenarios, on the other hand, SM fermions are centered in the brane at different locations and their mixings are due to overlapping wave functions. In order to reproduce observed neutrino mixings the relative locations of neutrinos in the brane are strictly constrained \cite{dienessarcevic}. However, incorporating bulk neutrinos to the split fermion models is a way to get round the fine tuning problems.      

Recently we have studied a Hybrid model of neutrinos \cite{dienes} from the viewpoint of leptogenesis \cite{ekapaperi}. The effective system consists of a Kaluza-Klein zero mode propagating in the bulk and two brane neutrinos.  The mass spectrum includes one light mass eigenstate that we identify as the electron neutrino of the Standard Model and two heavy states that are nearly degenerate. Our conclusion was that lepton and consequently baryon number are produced mainly via the mixing loop due to the heavy states. Our result for the CP violation parameter is more complicated with respect to earlier studies concerning the heavy particle mixing contribution to leptogenesis as the light neutrino has acquired a mass through the Higgs vacuum expectation value. 

Because of the complexity of the model in \cite{ekapaperi}, it is not easy to discern patterns on how CP violation depends on individual parameters while others are fixed. For this reason we elaborate on the allowed six parameters by performing a statistical analysis of the parameter space in the present paper. The allowed regions are determined by Monte Carlo analysis on the basis that the model consisting of two brane neutrino families generates sufficient CP violation for baryon number production and that the lightest neutrino mass remains within correct boundaries. We also pinpoint which higher order diagrams dominate in the CP violation parameter.

\section{The Model and CP Violation}
\setcounter{equation}{0}
\setcounter{footnote}{0}

The model consists of two brane neutrinos $\nu_{1,2}$ and one neutrino $\Psi=(\psi_+,\ol\psi_-)^T$ that lives in the bulk \cite{dienes, ekapaperi}. The brane neutrinos have an expression that is a Gaussian with respect to the brane location in the bulk $y_\alpha$
\beqa
\nu_\alpha(x,y)=\frac{1}{\sqrt{\sigma}}\exp\left(-\frac{\pi}{2}
\frac{(y-y_\alpha)^2}{\sigma^2}\right)\nu_\alpha(x).
\eeqa
The extra dimension is assumed to possess orbifold compactification and thus the even modes obey $\psi_+(-y)=\psi_+(y)$ and the odd modes satisfy $\ol\psi_-(-y)=-\ol\psi_-(y)$. 
\beqa
\psi_+(x,y)&=&\frac{1}{\sqrt{2\pi R}}\psi^{(0)}_+(x)
+\frac{1}{\sqrt{\pi R}}\sum_{n>0}\psi^{(n)}_+(x)\cos\frac{ny}{R} , \nn\\
\ol\psi_-(x,y)&=&
\frac{1}{\sqrt{\pi R}}\sum_{n>0}\ol\psi^{(n)}_-(x)\sin\frac{ny}{R}.
\eeqa

The five dimensional brane-bulk action reads 
\beqa \label{branebulkaction}
\mc S_c&=&\int d^4xdy\sum_{\alpha=1}^{n_f}
\Big\{M_*\nu^\dagger_\alpha(x,y)\left[\psi_+^c(x,y)
+e^{i\delta_\alpha}\ol\psi_-(x,y)\right]\\
%&&+\nu_\alpha ^\dagger(x,y)\frac{g}{\sqrt{2\pi R}}h(x)
&&+\nu_\alpha ^\dagger(x,y)g\, h(x,y)
\left[\psi_+^c(x,y)+e^{i\delta_\alpha}\ol\psi_-(x,y)\right]\Big\}
+\textrm{h.c.}\nn.
\eeqa
which after integrating over the extra dimension $y$ becomes
\beqa \label{action}
S_c&=&\int d^4x \sum_{\alpha=1}^{n_f}\Big\{\nu^\dagger_\alpha(x)\left [ m\psi^{(0)c}_+(x)+\sum_{n>0}\left ( m^\alpha_{n,+}\psi_+^{(n)c}(x)+m^\alpha_{n,-}\ol\psi^{(n)}_-(x)\right )\right ]\\
&&+\nu^\dagger_\alpha(x)\left [ \frac{hm}{v}\psi^{(0)c}_+(x)+\sum_{n>0}\left ( \frac{h(x)m^\alpha_{n,+}}{v}\psi_+^{(n)c}(x)+\frac{h(x)m^\alpha_{n,-}}{v}\ol\psi^{(n)}_-(x)\right )\right ]\Big\}+\textrm{h.c.},\nn
\eeqa
where 
\beqa \label{couplings}
m&\equiv &M_*\sqrt{\frac{\sigma}{\pi R}}=\frac{gv}{\sqrt{2 \pi R}}\sqrt{\frac{\sigma}{\pi R}},\nn \\
m^ \alpha _{n,+} &\equiv &\sqrt{2} m\cos \left(\frac{ny_\alpha}{R}\right)\textrm{exp}\left [-\frac{n^2\sigma^2}{2\pi R^2}\right ], \nn \\
m^ \alpha _{n,-} &\equiv &\sqrt{2} m e^{i\delta _\alpha}\sin\left(\frac{ny_\alpha}{R}\right)\textrm{exp}\left [-\frac{n^2\sigma^2}{2\pi R^2}\right ]
\eeqa
After integrating out the heavy Kaluza-Klein modes we are left with the effective 3$\times$3 mass matrix
\beqa \label{blockmass}
\widetilde{\mc M}_\textrm{L}&=&\mc M_\textrm{L}-\kappa \mc M^\textrm{T}_\textrm{D}=\begin{pmatrix} -\sum_n\big (m^\alpha_{n,-}m^\beta_{n,+}+m^\alpha_{n,+}m^\beta_{n,-}\big)\frac{R}{n} & m\\
m & 0 \end{pmatrix}\\
&\equiv & \begin{pmatrix} (m_{\alpha\beta}) & m\\
m^T & 0 \end{pmatrix}\nn,
\eeqa
where
\beqa
m_{\alpha\beta}&=&-\frac{M_*^2\sigma}{2}\Big\{e^{i\delta_\beta}\bigg[\Erf\big(\frac{\sqrt{\pi}}{2\sigma}(y_\alpha+y_\beta)\big)-\Erf\big(\frac{\sqrt{\pi}}{2\sigma}(y_\alpha-y_\beta)\big)\bigg]\\
&&+e^{i\delta_\alpha}\bigg[\Erf\big(\frac{\sqrt{\pi}}{2\sigma}(y_\alpha+y_\beta)\big)+\Erf\big(\frac{\sqrt{\pi}}{2\sigma}(y_\alpha-y_\beta)\big)\bigg]\Big\}.\nn
\eeqa
The physical mass eigenvalues obtained from (\ref{blockmass}) comprise one light state $m_1\simeq m_{\alpha\beta}$ and two heavier states with $m_{2,3}\simeq m$.

\begin{figure}[ht]
\centering
\subfigure[]{
\includegraphics[scale=1]{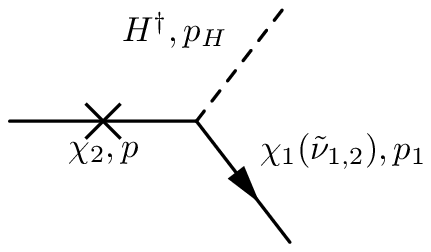}
\label{fig:subfig1}
}
\quad
\subfigure[]{
\includegraphics[scale=0.9]{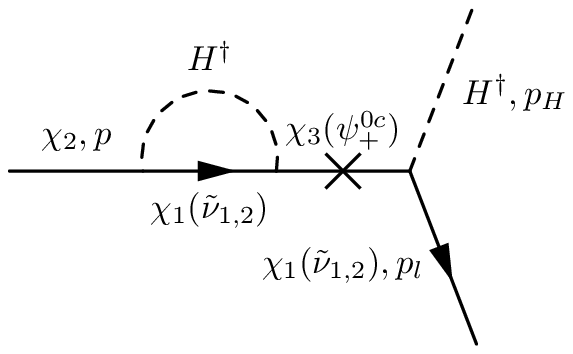}
\label{fig:subfig2}
}
\quad
\subfigure[]{
\includegraphics[scale=0.9]{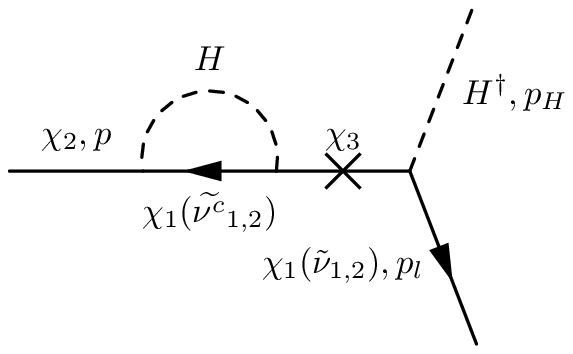}
\label{fig:subfig3}
}
\caption{The relevant Feynman diagrams for the process $\chi_2\rightarrow \chi_{1\textrm{L}} H^\dagger$ are shown. The tree level diagram due to the decay of $\chi_2$ to a neutrino and Higgs is in Fig \ref{fig:subfig1}. Fig \ref{fig:subfig2} and \ref{fig:subfig3} depict the mixing diagrams due to the the decay of $\chi_2$ to a light neutrino and Higgs.}
\label{fig:leptondecays}
\end{figure}

Lepton and baryon number are produced in the processes shown in Fig. \ref{fig:leptondecays} and the corresponding antineutrino production diagrams. Diagrams with mass insertion prior to the loop conserve total lepton number and vanish when summed over the final state neutrino flavors \cite{coviwf, leptogenesisreview}. Since the decaying heavy states are nearly degenerate, we expect the CP asymmetry to arise from the mixing diagrams shown.

%\begin{figure}[ht]
%\centering
%\subfigure[The tree level diagram due to the decay of $\chi_2$ to a light antineutrinoneutrino and Higgs.]{
%\includegraphics[scale=1]{ntoantinuhtreelevel.eps}
%\label{fig:antisubfig1}
%}
%\quad
%\subfigure[The mixing diagram due to the the decay of $\chi_2$ to a light antineutrino and Higgs. The mass insertion occurs prior to the loop.]{
%\includegraphics[scale=0.9]{ntoantinuhfirstmixingloop.eps}
%\label{fig:antisubfig2}
%}
%\quad
%\subfigure[The mixing diagram due to the the decay of $\chi_2$ to a light antineutrino and Higgs. The mass insertion occurs after the the loop.]{
%\includegraphics[scale=0.9]{ntoantinuhsecondmixingloop.eps}
%\label{fig:antisubfig3}
%}
%\caption{The relevant Feynman diagrams for the process $\chi_2\rightarrow \chi_{1\textrm{R}} H$.}
%\label{fig:antileptondecays}
%\end{figure}

We computed CP violation parameter in \cite{ekapaperi} and our result is 
\beqa \label{cpv}
\eps&=&a\cos\frac{\theta_2-\theta_3}{2}+b\sin \frac{\theta_2-\theta_3}{2},\\
a&=&\frac{1}{2}(m_2+m_3)^{-1}\Bigg \{m_2\Big[(m_2^2-m_3^2-|A_{33}|^2m_2^2)^2+4 m_2^4(\Rea A_{33})^2\Big]^{-1}\times \nn\\
&&\Big[(m_2^2-m_3^2-|A_{33}|^2m_2^2)(m_2m_3(\Ima A_{33}-\Ima A_{22})\nn\\
&&+m_2^2(\Rea A_{33}\Rea A_{22}+\Ima A_{33}\Ima A_{22}-|A_{33}|^2+\Ima A_{22}-\Ima A_{33}))\nn\\
&&+2 m_2^2\Rea A_{33}(m_2^2(-\Ima A_{33}\Rea A_{22}+\Rea A_{33}\Ima A_{22}-\Rea A_{22}+\Rea A_{33})\nn\\
&&+m_2m_3(\Rea A_{33}-\Rea A_{22}))\Big]\nn\\
&&+m_3\Big[(m_3^2-m_2^2-|A_{22}|^2m_3^2)^2+4m_3^4(\Rea A_{22})^2\Big]^{-1}\times\nn\\
&&\Big[(m_3^2-m_2^2-|A_{22}|^2m_3^2)(m_2m_3(\Ima A_{22}-\Ima A_{33})\nn\\
&&+m_3^2(\Rea A_{22}\Rea A_{33}+\Ima A_{22}\Ima A_{33}-|A_{22}|^2+\Ima A_{33}-\Ima A_{22}))\nn\\
&&+2m_3^2\Rea A_{22}(m_2m_3(\Rea A_{22}-\Rea A_{33})\nn\\
&&+m_3^2(-\Rea A_{33}\Ima A_{22}+\Rea A_{22}\Ima A_{33}-\Rea A_{33}+\Rea A_{22}))\Big]\Bigg\},\nn
\eeqa
where
\beqa
A_{33}&=&\frac{1}{256\pi}\frac{m^2}{v^2}\Big (\frac{m_{11}-m_{22}}{|m_{11}-m_{22}|}\Big )^2\big(e^{i(\theta_1-\theta_3)}+e^{i(\theta_1/2+\theta_2/2-\theta_3)}\big)(=A_{23}),\\
A_{22}&=&\frac{1}{256\pi}\frac{m^2}{v^2}\Big (\frac{m_{11}-m_{22}}{|m_{11}-m_{22}|}\Big )^2\big (e^{i(\theta_1-\theta_2/2-\theta_3/2)}
+e^{i(\theta_1-\theta_3)/2}\big )(=A_{32}).  \nn
\eeqa
The factor $b$ is similar in form to $a$. The contribution of the term proportional to $\sin ((\theta_2-\theta_3)/2)$ is negligible because the difference $\theta_2-\theta_3$ is suppressed by the large scale $m$ and thus the values of the sine function become small. Leaving this part out does not affect the allowed parameter regions. 

This result for $\eps$ deviates from the findings of earlier studies for heavy particle mixing and consequent resonant leptogenesis \cite{coviwf, liusegre, sarkarwf, pilaftsiscp}. For one thing, the couplings $A_{ij}$ do not satisfy $A_{ij}=A_{ji}^*$ because of the decompositions of the mass eigenstates. The light brane neutrino has acquired a mass and thus mixes with the other brane neutrino and KK zero mode in the mass matrix. In contrast, $SO(10)$ motivated models SM have massless leptons at the energy scale where the heavy neutrinos decay and thus the neutrino mass spectrum consists only of heavy states.

In the next section, we move forward to perform the numerical analysis which contains several physical constraints. First, there is the observed baryon number of the Universe which can be derived from the CP violation parameter. Also, the processes producing CP violation must have rates smaller than the expansion rate of the Universe. Finally, the light neutrino mass must obey certain limits.

%%%%%%%%%%%%%%%%%%%%%%%%%%%%%%%%%
\section{Numerical Analysis}
\setcounter{equation}{0}
\setcounter{footnote}{0}

We perform Monte Carlo analyses where we randomize over a set of values for 
\nl $(R,\tilde y_1,\tilde y_2, \delta_1, \delta_2, M_*^2\sigma)$. The physical constraints consist of the CP violation bound 
\newline $-1.6\times 10^{-6}< \eps < -1.6\times 10^{-7}$, the condition that the heavy neutrino decay rate must be less than the expansion rate of the Universe, $\Gamma\lesssim H$ and the experimental bound on the electron neutrino mass $m_{\nu_e}\lesssim 2.2$ eV \cite{lightneutrinomass}. The washout factor $\kappa$ in $Y_L=\kappa \eps /g_*$ is estimated to be 0.01-0.1 \cite{Hubble rate}-\cite{leptogenesislecture}, which gives the above mentioned constraint $-1.6\times 10^{-6}< \eps < -1.6\times 10^{-7}$ for the CP violation parameter. We have run the program so that there are around 20000 points for each parameter that satisfy the above conditions so in each of the plots there are around 20000 points. In the rest of this section, we present plots of various parameter planes where the allowed regions illustrate the essential features of the model.

First, we have made a scatter plot in the $(R,m_1)$ plane, $m_1$ being the light mass, shown in Fig. \ref{fig:subfig1}. The size of the extra dimension $R$ is allowed to vary in the interval $(10^{-17},10^{-7})\textrm{TeV}^{-1}$ and applying the conditions mentioned in the previous paragraph allow for $R$ values $(10^{-15},10^{-9})\textrm{TeV}^{-1}$. There are distinct bands that are favored, namely $R\simeq 10^{-13}$ TeV$^{-1}$ and $R\simeq 10^{-11}$ TeV$^{-1}$. A similar pattern is produced in the $(R,\tilde y_{1,2})$ planes with the same favored values of $R$.

\begin{figure}[ht]
\centering
\subfigure[]{
\includegraphics[scale=0.6]{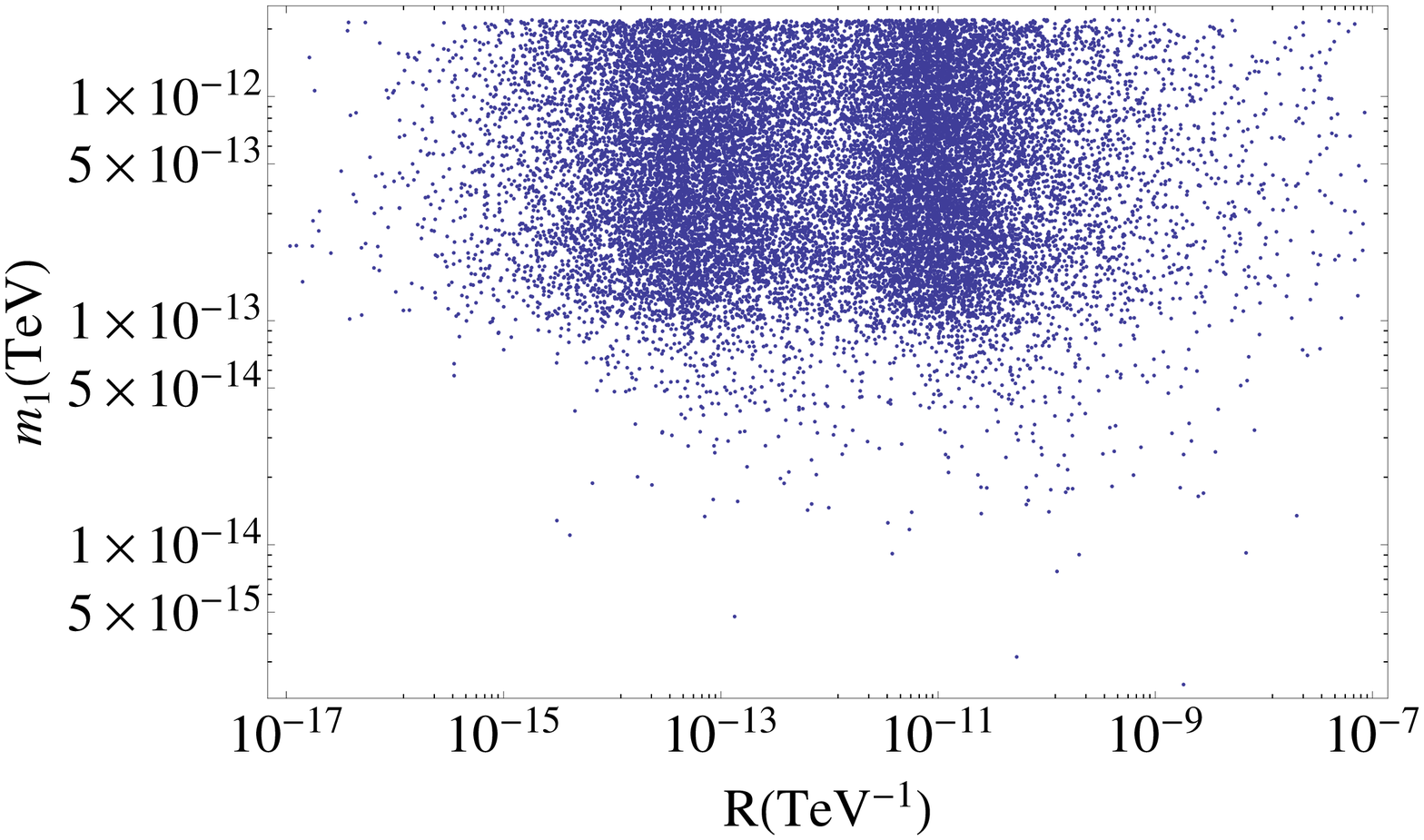}
\label{fig:subfig4}
} 
\quad
\subfigure[]{
\includegraphics[scale=0.6]{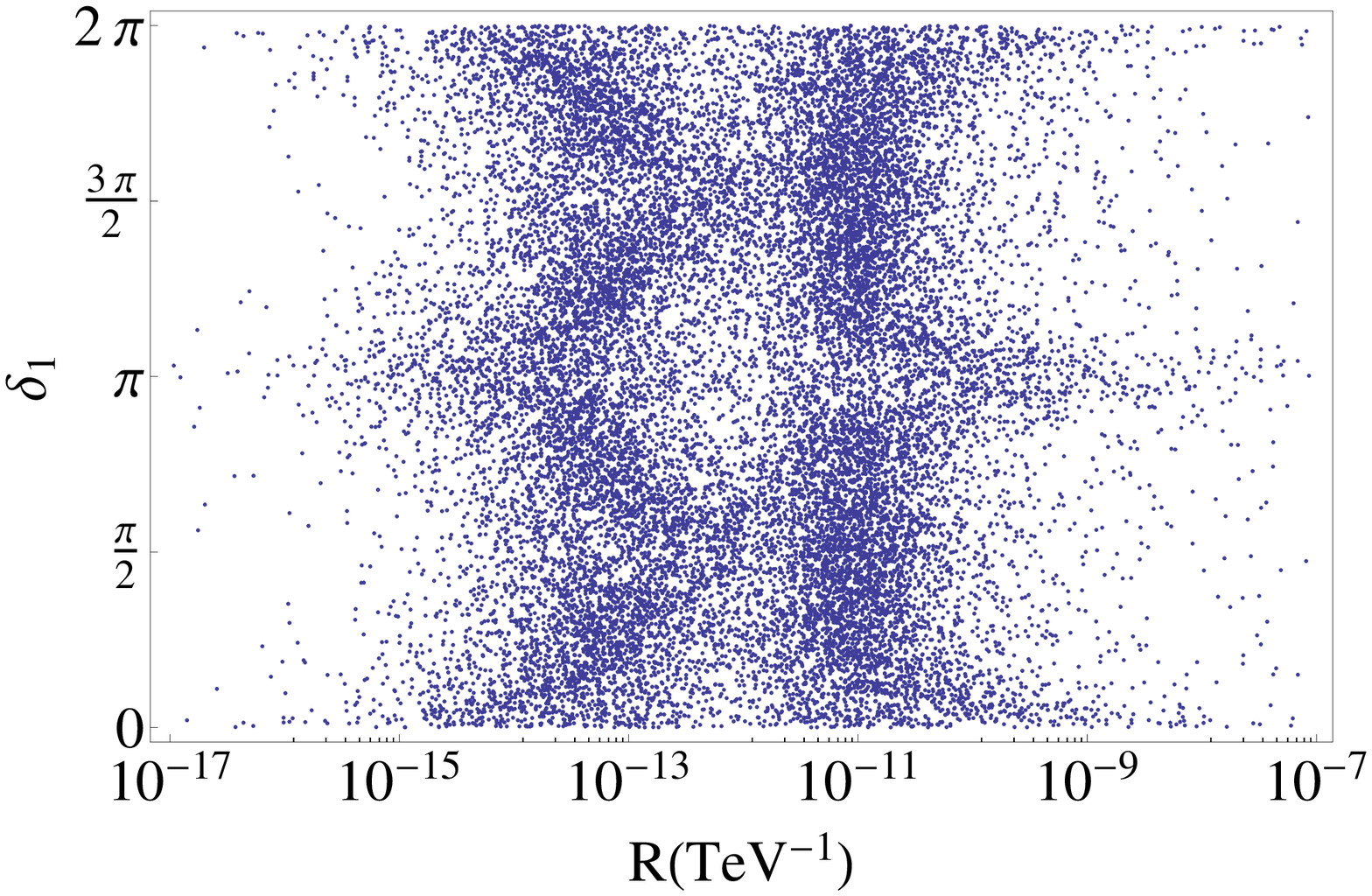}
\label{fig:subfig5}
}
\caption{Fig. \ref{fig:subfig4} shows the scatter plot between $R$ and $m_1$. Clearly, $R\simeq 10^{-11}$ TeV$^{-1}$ and $R\simeq 10^{-13}$ TeV$^{-1}$ are favored. Fig. \ref{fig:subfig5} depicts the scatter plot between $R$ and $\delta_1$.}
\label{fig:Rmd}
\end{figure}

%Fig \ref{ekaplot} depicts the relation between $R$ and $\tilde y_1$ and the dependence between $R$ and $\tilde y_2$ is analogous.

%\begin{figure}[ht]
%\centering
%\includegraphics[scale=0.80]{MCRvsy1500k091123.eps}
%\caption{A scatter plot between $R$ and $\tilde y_1$. Clearly, $R\simeq 10^{-11}$ TeV$^{-1}$ and $R\simeq 10^{-13}$ TeV$^{-1}$ are favored. Also, the size of the extra dimension is restricted to close the Planck scale.}
%\label{ekaplot}
%\end{figure} 

In the $(R,\delta_{1,2})$ planes, see Fig. \ref{fig:subfig5} $(R,\delta_1)$ plane, similar patterns are observed as in $(R,m_1)$ and $(R,\tilde y_{1,2})$ planes with the distinction that now there is more variation in the $R$ values which is reflected in the zigzag-like pattern. Clearly, at $\delta_{1,2}\simeq n\pi$ the interval $10^{-13}\textrm{TeV}^{-1}\lesssim R\lesssim 10^{-11}\textrm{TeV}^{-1}$ is disfavored over small $R< 10^{-13}\textrm{TeV}^{-1}$ and large $R>10^{-11}\textrm{TeV}^{-1}$ values. At $\delta_{1,2}\simeq n\pi /2$ the situation reverses and the interval $10^{-13}\textrm{TeV}^{-1}\lesssim R\lesssim 10^{-11}\textrm{TeV}^{-1}$ is favored over extreme $R$ values.

When we consider the dependence between the light and heavy masses, we yet obtain a figure with two distinct bands that are slightly tilted compared to the ones in the $(R,m_1)$ plane. This graph bears a visual resemblance to Fig. \ref{fig:subfig4} since the heavy masses $m_{2,3}\sim R^{-1/2}$. A similar dependence is found between $m_3$ and $m_1$. 

%\begin{figure}[ht]
%\centering
%\includegraphics[scale=0.70]{MCm2vsm1500k091126.eps}
%\caption{A scatter plot between $m_2$ and $m_1$. The mass of the heavy decaying state resides at the TeV scale.}
%\label{kolmasplot}
%\end{figure}

Fig. \ref{fig:subfig4} showcases a distinct pattern where two regions of $R$ values are favored, namely $R\sim 10^{-13}$TeV$^{-1}$ and $R\sim 10^{-11}$TeV$^{-1}$. Similar behaviour is presented in Fig. \ref{fig:subfig5} where the two bands form a zigzag-like pattern suggesting that $R$ depends on the phase angles $\delta_{1,2}$. When all parameters save $R$ are fixed, we make the following observations as the phase angles $\delta_{1,2}$ are varied. 

The larger values $R\sim 10^{-11}$TeV$^{-1}$ are reached when the phase angles $\delta_{1,2}$ lie in the I quadrant and are nearly degenerate, {\it{e.g.}} $\delta_1=\pi/12$ and $\delta_2=\pi/10$. When the values are further apart (still both in the I quadrant), the size of the extra dimension $R$ has to be of the order $(10^{-13}-10^{-12})$TeV$^{-1}$. 
If both phase angles $\delta_{1,2}$ take values in the second and third quadrants, the allowed region for $R$ lies in the vicinity of $R\sim 10^{-13}$TeV$^{-1}$. As the angles move to IV quadrant, larger $R$ values $\sim 10^{-11}$TeV$^{-1}$ are again favored. If the angles lie in different quadrants, then for instance $\delta_1 \in [0,\pi/2)$ and $\delta_2 \in (\pi/2,\pi)$ require $R\sim 10^{-13}$TeV$^{-1}$. The same applies if $\delta_1 \in [0,\pi/2)$ and $\delta_2 \in (\pi,3 \pi/2)$. As $\delta_2$ progresses to the IV quadrant, the larger end of $R$ values are preferred. The larger end of $R$ values is also required in the cases where $\delta_1 \in (\pi/2,\pi)$ and $\delta_2 \in (3 \pi/2, 2 \pi)$ as well as with $\delta_1 \in (\pi, 3\pi/2)$ and $\delta_2\in (3 \pi/2,2\pi)$. 

\begin{figure}[ht]
\centering
\includegraphics[scale=1]{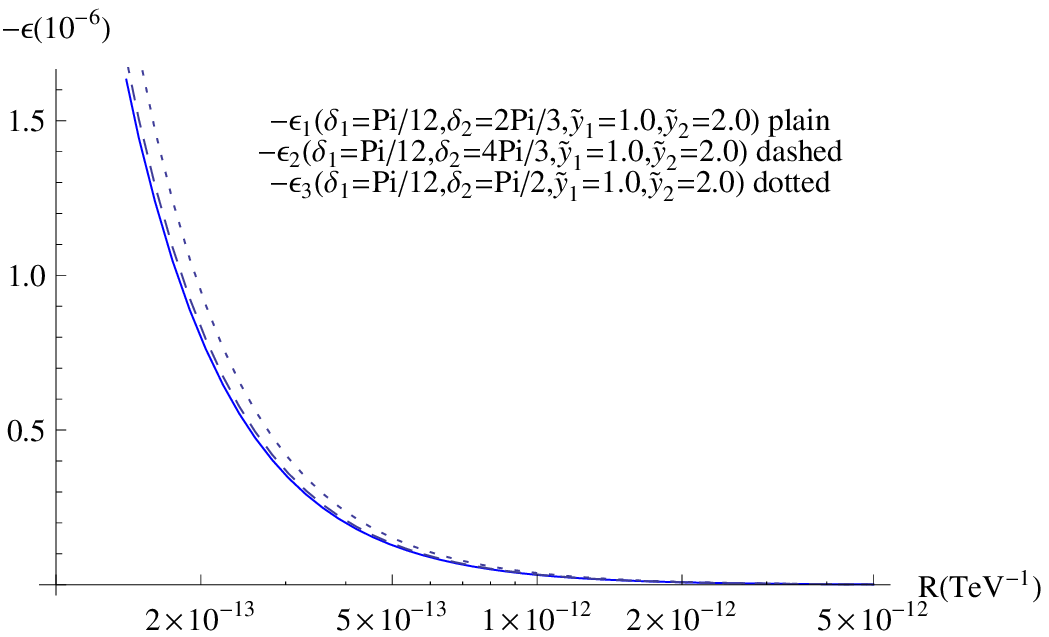}
\caption{The CP violation parameter $\eps$ as a function of the size of the extra dimension $R$, with $-\eps_1(\tilde y_1=1.0$,$\tilde y_2=2.0$,$\delta_1=\pi/12$,$\delta_2=2\pi/3)$,  $-\eps_2(\tilde y_1=1.0$,$\tilde y_2=2.0$,$\delta_1=\pi/12$,$\delta_2=4\pi/3)$,  $-\eps_3(\tilde y_1=1.0$,$\tilde y_2=2.0$,$\delta_1=\pi/12$,$\delta_2=\pi/2)$.}
\label{Rvseps}
\end{figure}

Fig. \ref{Rvseps} illustrates some of the behavior of $\eps$ as a function of $R$ when the phase angles $\delta_{1,2}$ are varied. The solid curve in Fig. \ref{Rvseps} corresponds to the set $\tilde y_1=1.0$, $\tilde y_2=2.0$, $\delta_1=\pi/12$ and $\delta_2=2\pi/3$, the dashed curve to the set $\tilde y_1=1.0$, $\tilde y_2=2.0$, $\delta_1=\pi/12$ and $\delta_2=4\pi/3$, and the dotted curve to the set $\tilde y_1=1.0$, $\tilde y_2=2.0$, $\delta_1=\pi/12$ and $\delta_2=\pi/2$.

\begin{figure}[ht]
\centering
\subfigure[]{
\includegraphics[scale=0.6]{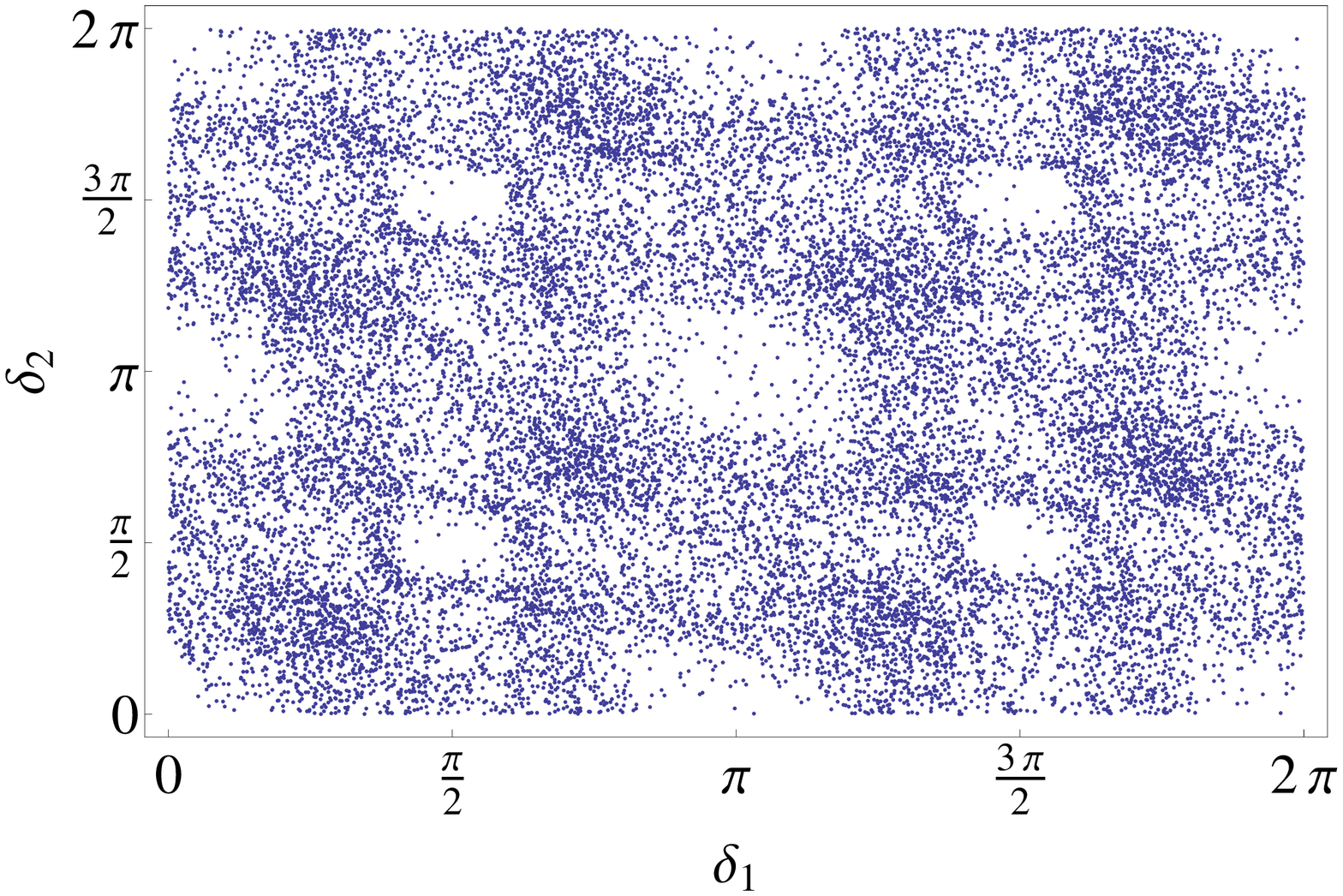}
\label{fig:fsubfig6}
}
\quad
\subfigure[]{
\includegraphics[scale=0.6]{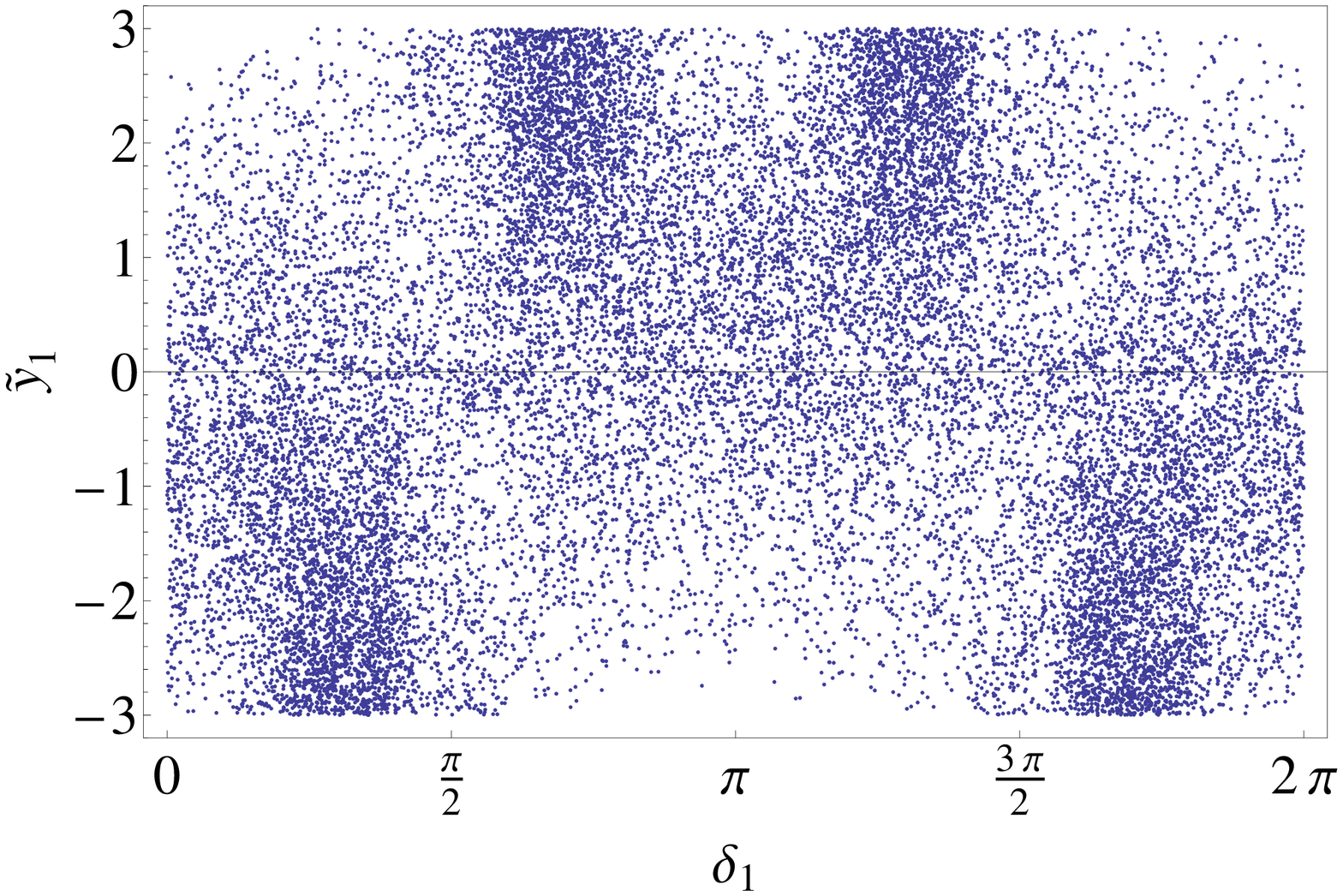}
\label{fig:fsubfig7}
}
\caption{Fig. \ref{fig:fsubfig6} shows the scatter plot between $\delta_1$ and $\delta_2$, where the size of the extra dimension is fixed, $R\sim 10^{-13}$TeV$^{-1}$. The scatter plot between $\delta_1$ and $\tilde y_1$ is found in Fig. \ref{fig:fsubfig7}, also with the same fixed $R$.}
\label{fig:fixedRfigs}
\end{figure}

Since two values  of $R$ clearly stand out in Fig. \ref{fig:Rmd}, we have also studied the behaviour of the system when $R$ is fixed to the most favored values $R\lesssim 10^{-13}$TeV$^{-1}$ and $R\lesssim 10^{-11}$TeV$^{-1}$. The smaller $R$ value produces Fig. \ref{fig:fsubfig6} which shows how the CP phases $\delta_1$ and $\delta_2$ are restricted. When both phases equal a multiple of $\pi$ or $\pi/2$, there is a void which indicates these values do not produce the correct magnitude for $\eps$. This can also be found out when the light neutrino mass $m_1$ is plotted against the phase angle $\delta_1$ or $\delta_2$, resulting in four distinct bands away from $\delta_{1,2}\simeq k\pi/2$ ($k=1,2,3,...$). When $R\sim 10^{-11}$TeV$^{-1}$, the voids at half-integer phase angle values $(\delta_1\simeq n\pi/2,\delta_2\simeq k\pi/2)$ are present neither in the $(\delta_1,\delta_2)$ plot of Fig. \ref{fig:fsubfig6} nor the $(m_1,\delta_{1,2})$ plots. 

The fixing of $R$ influences the relationship between $\tilde y_{1,2}$ and $\delta_{1,2}$, as well. In Fig. \ref{fig:fsubfig7} we see how the phase $\delta_1$ and the brane location $\tilde y_1$ depend on each other at $R\sim 10^{-13}$TeV$^{-1}$. Integer and half-integer multiples of $\pi$ are less favored overall, and in the I and IV quadrants negative $\tilde y_1$ values are favored and in the II and III quadrants positive $\tilde y_1$ values are favored. A similar pattern is found for $\delta_2,\tilde y_2)$.  Plots confronting $(\delta_2,\tilde y_1)$ and $(\delta_1,\tilde y_2)$ show continuous bands over the $\tilde y_{1,2}$ ranges and less points at $\delta_{1,2}\simeq k\pi/2,\ k=1,2,3,...$. If $R\sim 10^{-11}$TeV$^{-1}$, the half-integer multiples of $\pi$ do not appear as unfavored regions in the above mentioned scatter plots.

The comparison of various terms in (\ref{cpv}) shows that the largest contribution to $\eps$ arises from terms $\sim m^4A_{ij}^2$ in the numerator as the linear terms $\sim m^4A_{ij}$ undergo cancellation and yield a factor that is many orders of magnitude smaller than $m^4A_{ij}^2$. Thus, in contrast to earlier studies of resonant leptogenesis \cite{pilaftsiscp}, where the higher order Yukawa terms  $\sim A_{ij}^2$ were found to cancel, it is crucial that in our case the higher order Yukawa terms are taken into account, despite the fact that one would expect them to be unimportant due to their higher order in  the Yukawa coupling. Numerically they turn out to be of equal magnitude with the terms $\sim m^4A_{ij}\sim m^4(m/v)^2$ in our model at the allowed $R$ scale. Consequently, the main contribution to $\eps$ comes from two-loop diagrams which are reflected in the numerator terms that go as $\sim m^4A_{ij}^2\sim m^4(m/v)^4$.

\section{Conclusions}
\setcounter{equation}{0}
\setcounter{footnote}{0}

In summary, we conclude that the size of the extra dimension is restricted to a few orders of magnitude away from the Planck scale when the constraints from the observed baryon-antibaryon asymmetry and electron neutrino mass are applied and  the higher order Yukawa terms are taken into account. The heavy neutrino mass lies in the TeV scale. If there were more extra dimensions, the sizes of these dimensions could be larger. However, this interesting case would require a separate study as the hierarchy is likely to change completely w.r.t. the one extra dimension case \cite{dienes}. 

As our scenario consists, in addition to two heavy neutrinos, just one
light neutrino, it is effectively a one flavor model. Therefore it cannot
be confronted with phenomena involving  mixing among the three light
neutrino flavors $\nu_e,\nu_{\mu},\nu_{\tau}$, such as neutrino
oscillations and neutrinoless double beta decay. One could extend the
model to include three light neutrino flavors by adding two extra brane
neutrinos. In this case the mass spectrum would consist of of three light
and two heavier neutrinos. It would be subject of a further study to
figure out whether  the mixing among the light neutrinos can be realized
consistently with observations in such scenario. It should be noted
anyhow, that since the heavy neutrinos in such a scenario would be in a
super-eV scale and furthermore unstable, the scenario would not be a
viable candidate for a 3+2 model recently considered as a possible
solution the LNSD and MiniBoone anomalies \cite{sterilenuexperiment}-\cite{pdg}.

Our model differs from earlier studies \cite{sarkarwf,coviwf,pilaftsiscp} in that the light brane neutrino has a mass term, implying that  the mass eigenstates have also a light neutrino component. If only the terms $\mc O(A_{ij})$ are included, cancellations occur in the CP violation parameter $\eps$, resulting in a CP violation that is too small to explain the observed baryon-antibaryon asymmetry of the Universe. As a consequence, the couplings $A_{ij}$ are different in form from those in \cite{sarkarwf,coviwf,pilaftsiscp} and we do not observe the cancellation of higher order Yukawa terms $\sim A_{ij}^2$ in contrast with the earlier studies. Actually, the situation is quite the opposite as in our case the linear $A_{ij}$ terms nearly cancel. Thus, $\eps$ cannot be considered only as an expansion in terms of the Yukawa coupling $\sim m^2/v^2$ but also the relative magnitudes of $m^2/v^2\times m_{2,3}$ and $|m_2-m_3|$ have to be considered and higher order Yukawa terms need to be included for $\eps$ to be sufficiently large.
\bigbreak

{\bf Acknowledgments} The work of H.K. was supported by Finnish Academy of Science and Letters (the Väisälä Fund).


\begin{thebibliography}{99}
\bibitem{smasymmetry}M. S. Turner in {\textit{Intersection Between Particle Physics and Cosmology}}, Jerusalem Winter School for Theoretical Physics, Vol. 1, p. 99, eds T. Piran and S. Weinberg (World Scientific, Singapore, 1986).
\bibitem{wmap}WMAP Collaboration, C. L. Bennett, {\textit{et al}}, Astrophys. J. Suppl. 148:1 (2003); D. N. Spergel, {\textit{et al}}, Astrophys. J. Suppl. 148:175 (2003).
\bibitem{Fukugita:1986hr} M. Fukugita and T. Yanagida, {\textit{Baryogenesis Without Grand Unification}} Phys.\ Lett.\  {\bf B} 174:45 (1986).
\bibitem{luty}M. A. Luty, {\textit{Baryogenesis via Leptogenesis}}, {\textit{Phys. Rev.}} {\bf D}45:455-465 (1992).
\bibitem{krs} V. A. Kuzmin, V. A. Rubakov and M. A. Shaposhnikov, {\textit{On the Anomalous Electroweak Baryon Number Nonconservation in the Early Universe}}, Phys. Lett. {\bf B155}, 36 (1985).
\bibitem{mohapatra}P. Minkowski, {\textit{Mu $\to$ E Gamma At A Rate Of One Out Of 1-Billion Muon Decays?}},Phys. Lett. B 67, 421 (1977);
M. Gell-Mann, P. Ramond and R. Slansky, {\textit{Complex Spinors And Unified Theories}}, 
in {\it Supergravity}, edited by 
P. van Nieuwenhuizen and D. Freedman (North-Holland, Amsterdam, 1979),  
p. 315; T. Yanagida in {\it Proceedings of the Workshop on Unified Theory 
and Baryon Number in the Universe}, edited by O. Sawada and A.~Sugamoto 
(KEK, Tsukuba, Japan, 1979); S. L. Glashow, {\textit{The future of elementary particle physics}}, in {\it Proceedings of the 1979 Carg`ese
Summer Institute on Quarks and Leptons}, edited by M. L´evy, J.-L. Basdevant, D. Speiser,
J.Weyers, R. Gastmans, and M. Jacob, Plenum Press, New York, 1980, pp. 687–
713; R. N. Mohapatra, G. Senjanovic, {\textit{Neutrino Mass and Spontaneous Parity Nonconservation}}, Phys. Rev. Lett 44:912-915 (1980).
\bibitem{arkanihamedrussell}N. Arkani-Hamed, S. Dimopoulos, G.R. Dvali, J. March-Russell, {\textit{Neutrino masses from large extra dimensions}}, Phys. Rev. {\bf D}65:024032 (2001).
\bibitem{dienesdudasgherghetta}K. R. Dienes, E. Dudas, T. Gherghetta, {\textit{Neutrino oscillations without neutrino masses or heavy mass scales: A Higher dimensional seesaw mechanism}}, Nucl. Phys. {\bf B}557:25-59 (1999).
\bibitem{arkanihamed}N. Arkani-Hamed, S. Dimopoulos, G. R. Dvali, {\textit{The Hierarchy problem and new dimensions at a millimeter}}, Phys. Lett. {\bf B}429:263-272 (1998); N. Arkani-Hamed, S. Dimopoulos, G. R. Dvali, {\textit{ Phenomenology, astrophysics and cosmology of theories with submillimeter dimensions and TeV scale quantum gravity}}, Phys. Rev. {\bf D}59:086004 (1999); I. Antoniadis, N. Arkani-Hamed, S. Dimopoulos, G. R. Dvali, {\textit{New dimensions at a millimeter to a Fermi and superstrings at a TeV}}, Phys. Lett. {\bf B}436:257-263 (1998).
\bibitem{dienes}Keith R. Dienes, Sabine Hossenfelder, {\textit{Hybrid model of neutrino masses and oscillations: Bulk neutrinos in the split-fermion scenario}}, Phys. Rev. {\bf D}74:065013 (2006).
\bibitem{thickbrane}N. Arkani-Hamed, M. Schmaltz, {\textit{Hierarchies without symmetries from extra dimensions}}, Phys. Rev. {\bf D}61, 033005 (2000); E. A. Mirabelli, M. Schmaltz, {\textit{Yukawa hierarchies from split fermions in extra dimensions}}, Phys. Rev. {\bf D}61, 113011 (2000).
\bibitem{dienessarcevic}K. R. Dienes, I. Sarcevic, {\textit{Neutrino flavor oscillations without flavor mixing angles}} Phys. Lett. {\bf B}500:133-141 (2001).
\bibitem{ekapaperi} J. Maalampi, I. Vilja, H. Virtanen, {\textit{Thermal leptogenesis in a 5D split fermion scenario with bulk neutrinos}}, Phys. Rev. D82:013009 (2010).
\bibitem{coviwf}L. Covi, E. Roulet, {\textit{Baryogenesis from mixed particle decays}}, Phys. Lett. {\bf B}399:113-118 (1997).
\bibitem{leptogenesisreview}S. Davidson, E. Nardi, Y. Nir, {\textit{Leptogenesis}}, Phys. Rept. {\bf 466}:105 (2008).
\bibitem{liusegre}J. Liu, G. Segré, {\textit{Reexamination of generation of baryon and lepton asymmetries in the early Universe by heavy particle decay}}, Phys. Rev. {\bf D}48:4609-4612 (1993).
\bibitem{sarkarwf}M. Flanz, E. A. Paschos, U. sarkar, J. Weiss, {\textit{Baryogenesis through mixing of heavy Majorana neutrinos}}, Phys. Lett. {\bf B}389:693-699 (1996).
\bibitem{pilaftsiscp}A. Pilaftsis, {\textit{CP violation and baryogenesis due to heavy Majorana neutrinos}}, Phys. Rev. {\bf D}56:5431-5451 (1997); A. Pilaftsis, {\textit{Heavy Majorana neutrinos and baryogenesis}}, Int. J. Mod. Phys. A14:1811-1858 (1999); A. Pilaftsis, T. E. J. Underwood, {\textit{Resonant leptogenesis}}, Nucl.Phys.B692:303-345 (2004).  
\bibitem{lightneutrinomass}J. Bonn et al., {\textit{The Mainz neutrino mass experiment}}, Nucl. Phys. B(Proc. Suppl.) 91,273 (2001).
\bibitem{Hubble rate} E.W. Kolb, M. S. Turner, {\textit{The Early Universe}}, Addison-Wesley Publishing Company (1993).
\bibitem{kappafactor}W. Buchmüller, T. Yanagida, {\textit{Quark lepton mass hierarchies and the baryon asymmetry}}, Phys. Lett. {\bf B}445:399-402 (1999).
\bibitem{leptogenesislecture}M.-C. Chen, {\textit{TASI 2006 Lectures on Leptogenesis}}, hep-ph/0703087 (2007).
\bibitem{sterilenuexperiment}M. Maltoni, T. Schwetz, {\textit{Sterile neutrino oscillations after first MiniBooNE results}}, Phys. Rev. D{\bf 76}:093005 (2007).
\bibitem{cosmologysternus}J. Hamann, S. Hannestad, G. G. Raffelt, I. Tamborra and Y. Y. Y. Wong, {\textit{Cosmology favoring Extra Radiation and Sub-eV Sterile Neutrinos as an Option}}, Phys. Rev. Lett. {\bf 105}:181301 (2010).
\bibitem{pdg}K. Nakamura {\it et al.} (Particle Data Group), J. Phys. G {\bf 37}:075021 (2010) and references therein.

\end{thebibliography}
\end{document}